\documentstyle[12pt]{article}
\begin{document}
\title{A Diffusion Equation for Quantum Adiabatic Systems }
\author{ Sudhir R. Jain \\
Theoretical Physics Division, Bhabha Atomic Research Centre \\
Trombay, Mumbai 400 085, India }
\date{}
\maketitle

\begin{abstract}
For ergodic adiabatic quantum systems, we study the evolution of energy
distribution as the system evolves in time. Starting from the von Neumann
equation for the density operator, we obtain the quantum analogue of the
Smoluchowski equation on coarse-graining over the energy spectrum. This
result brings out the precise notion of quantum diffusion.

\end{abstract}
\newpage

There are many physical situations where a separation of time scales occurs,
the adiabatic approximation to an evolution is one of the expressions of
such a separation \cite{ll}. Classically, an important problem in this
context is to describe the evolution in phase space of an ensemble of
systems under an ergodic adiabatic Hamiltonian \cite{cj}. An ergodic
adiabatic Hamiltonian can be defined as one which describes ergodic
(usually specialized to chaotic) dynamics under a slowly time-dependent
Hamiltonian. In this paper, we study the quantal version of this treatment
and obtain results which, in an appropriate limit, reduce to results in
\cite{cj}. The reasons to present quantum analogue of the classical case
study are many-folded. Firstly, classical physics being a special case of
quantum physics, it is interesting to know what is the equation that one
gets so that the limiting situation is clearly interpreted. Secondly,
multiple time-scale analysis is usually employed on
classical equations and it is, therefore, interesting to expose how this
method may be shown to be fruitful  for  quantal equations also
\cite{bender}.  Thirdly, a clear analysis based on time-scale separation
of the von Neumann equation paves way for the next-order
complication that arises in combining the adiabatic and semiclassical
limits \cite{sister}. The motivation behind this combination of singular
limits is provided by a work where first-order velocity-dependent
corrections to the lowest adiabatic approximation for the reaction
force on the slow system are studied \cite{br}. In the classical setting,
one recovers geometric magnetism and deterministic friction as the reaction
forces whereas in the half-classical setting (fast system treated quantum
mechanically) there is only geometric magnetism, no friction. Deterministic
frictional force was found in \cite{mw} and is non-zero when the fast motion
is classical and chaotic.

Very interesting examples can be cited from different fields of physics and
chemistry that resonate with the abovementioned problem. One of the
well-studied problem is when there are many particles moving in a
time-dependent shape of the box \cite{cj-ws}. This is an idealization of
nuclear fission and fusion. It has been shown in a numerical experiment
that the transition from ordered to chaotic nucleonic motions is accompanied
by a transition in collective properties of nuclei from those of elastic
solid to visco-elastic to viscous fluid \cite{bss}.
Quantum mechanical origin of dissipation in finite Fermi systems has been
recently explained \cite{srj-akp}. It has been found that the geometric
phase, $\gamma $ acquired by a single-particle wavefunction in adiabatic perturbation
is related to the absorptive part of the frequency-dependent response
function, $\chi ^{\prime \prime}$ of the finite bulk, hence dissipation -
we call them $\gamma - \chi $ relations.
Mesoscopic systems can also show this kind of behaviour in their
conductivity properties.

We now describe the basic problem.
Consider a quantum system evolving adiabatically in time. Since the
evolution is adiabatic, we have energy levels at every instant of time
and an instantaneous basis. Let us now assume that the energy levels do
not cross and the reason for this be left unspecified. In particular,
this may arise in a random matrix hypothesis for the system. During the
evolution, although the levels do not cross, they may come arbitrarily
close to each other. Very small spacing between levels then leads to an
increased probability of non-adiabatic Landau-Zener transitions
\cite{zener} which  eventually change the energy distribution of the
system. The basic idea behind using the Landau-Zener transitions goes
back \cite{h-w} in the literature of nuclear physics where it is used
to explain  damping of collective modes.

The classical version of the problem treated here has a distinguished
history. Starting from the earlier works of Ehrenfest, in classical
context, it was shown first by Lenard \cite{lenard} that if Hamiltonian,
$H = $constant for $t < 0$ and $t > t_1$, then the values of the reduced
action $\oint ~pdq$ for $t < 0$ and $t > t_1$ differ from each other by
${\cal O}(\epsilon ^m)$, however large $m$ may be. Generality of the
adiabatic invariant led Ott \cite{ott} to show that the error in ergodic
adiabatic invariant in the classical version of our problem is diffusive,
however the equation differed from the Smoluchowski equation. The problem
was re-examined by Jarzynski \cite{cj} where the Smoluchowski equation was
restored (although the authors of \cite{ott,cj} call it by the name of the
Fokker-Planck equation). The quantum version is the subject here. To start
with, due to important differences between classical and quantum mechanics,
and with lack of a precise notion of chaos in quantum systems \cite{srjain},
it is not clear what will the final equation for energy distribution be.
We now show that the equation is different from the classical case.

To begin the analysis, let us consider the Hamiltonian,
\begin{equation}
\hat{H}(t) = \hat{H}_0 + \epsilon t~\hat{V}
\end{equation}
where $\hat{H}_0, \hat{V}$ are linear operators. We assume that the
evolution in time is adiabatic which corresponds to the smallness of
$\epsilon $.  As a concrete realization where avoided level crossings
may occur, one may assume that the linear operators $\hat{H}_0$ and
$\hat{V}$ belong to some random matrix ensembles invariant under a
canonical group where it may be mentioned that a detailed study on
time correlation functions already exists \cite{gjv}. Thence, at any
instant, system admits an eigenvalue spectrum given by the eigenvalue
problem for the ``frozen" Hamiltonian,
\begin{equation}
\hat{H}(\epsilon t)|n(\epsilon t)\rangle  = E_n(\epsilon t)|n(\epsilon t)\rangle .
\end{equation}
The probability of the system residing in the state $|n(\epsilon t)\rangle $ at
time $t$ is given in terms of the density operator, $\hat{\rho }$,
\begin{equation}
p_n(t) = \langle n(\epsilon t)|\hat{\rho }|n(\epsilon t)\rangle ,
\end{equation}
with
\begin{equation}
\sum_{n}~p_n(t) = 1 ~~\forall ~t.
\end{equation}
On the other hand, the probability of finding the system in the range
$\Delta E$ about energy $E$, is defined through
\begin{equation}
p(E)dE = \sum_{E \leq E_n \leq E+\Delta E}~p_n(t)
\end{equation}
with
\begin{equation}
\int_{-\infty }^{\infty } dEp(E) = 1.
\end{equation}
To relate $p_n$ and $p(E)$, let us define the density of states,
\begin{equation}
\Sigma (E) = \frac{\Delta N}{\Delta E}
\end{equation}
so that
\begin{equation}
p(E)\Delta E = p(E) \frac{\Delta N}{n(E)} = p(N)\Delta N
\end{equation}
implying thereby
\begin{equation}
p(E) = \Sigma (E)p(N).
\end{equation}
Eq. (9) relates the probability of finding N levels in the energy range
$[E,E+\Delta E]$ and the probability of finding the system in the state
corresponding to energy $E$.

From time $t=0$,the levels  evolve  in time, the resulting density
operator, $\hat{\rho }$ satisfies
\begin{equation}
i\hbar \frac{\partial \hat{\rho }}{\partial t} = [\hat{H},\hat{\rho }].
\end{equation}
Starting from (10), our objective is to derive an equation for the energy
distribution,
\begin{equation}
\eta (E) = \int^{E} dE'\mbox{tr}\{ \delta(E'-\hat{H})\hat{\rho } \}.
\end{equation}
Since $\epsilon $ in (1) is small (adiabaticity) parameter, we have
two time-scales - $t$ ("fast" scale) and $\epsilon t$ ("slow" scale).
To incorporate these scales in the problem, we employ the multiple
time-scale method for treating the partial differential equation (10).
Accordingly, denoting the set of instantaneous states by
$\{|n(\epsilon t)\rangle \}$, we can write an expansion for the
density operator,
\begin{equation}
\hat{\rho }(\{|n(\epsilon t)\rangle \},t) = \hat{\rho }_0
(\{|n(\epsilon t)\rangle \},\epsilon t) + \epsilon \hat{\rho }_1
(\{|n(\epsilon t)\rangle \},t,\epsilon t) + ...
\end{equation}
with the initial conditions,
\begin{equation}
\hat{\rho }_0 = \hat{\rho }_{00}(\hat{H}(\epsilon t)),~\hat{\rho }_1=\hat{\rho }_2=\hat{\rho }_3=...=0.
\end{equation}
Substituting (12) in (10), we get a system of equations separated by
different orders of $\epsilon $ :
\begin{eqnarray}
[\hat{\rho }_0,\hat{H}(\epsilon t)] &=& 0, \\
i\hbar \frac{\partial \hat{\rho }_j}{\partial t} +
[\hat{\rho }_j,\hat{H}(\epsilon t)] &=& -i\hbar \frac{\partial
\hat{\rho }_{j-1}}{\partial \epsilon t}, ~~j=1,2,...
\end{eqnarray}
If there are no other constants of the motion than $H(\epsilon t)$ on
the fast scale, or under the Thomas-Fermi approximation, by (14),
\begin{equation}
\hat{\rho }_0(\{|n(\epsilon t)\rangle \},\epsilon t) =
\hat{\rho }_0'(H(\epsilon t),\epsilon t)
\end{equation}
where the arbitrariness of $\hat{\rho }_0'$ is removed by insisting
that $\hat{\rho }$ remains valid for times $O(\epsilon ^{-1})$ by
removing secularities in (15) with $j=1$. To realise this, we operate
on the $j=1$ equation by an arbitrary operator-valued function \cite{foot1},
$g(\hat{H})$ and perform the trace of the resulting equation over
the frozen basis,
\begin{equation}
\sum_{n}\left< n\vline g\frac{\partial \hat{\rho }_1}{\partial t}
\vline n\right>  + \frac{1}{i\hbar }\sum_{n} \langle n|g[\hat{\rho }_1,
\hat{H}]|n\rangle = - \sum_{n}\left< n\vline g
\frac{\partial \hat{\rho }_0}{\partial (\epsilon t)} \vline n\right> .
\end{equation}

For $\hat{\rho }_0$ to be valid for times for $O(\epsilon ^{-1})$,
the right hand side (RHS) of (17) should be set to zero, which leads to

\begin{equation}
\sum_{n}\left< n\vline g
\left[ \frac{\partial \rho _0'(\hat{H})}{\partial E}
\frac{\partial \hat{H}}{\partial (\epsilon t)} +
\frac{\partial \rho _0'(\hat{H})}{\partial (\epsilon t)} \right]
\vline n\right> = 0.
\end{equation}
Let us now define
\begin{eqnarray}
\Sigma (E,\epsilon t) &:=& \sum_{n} \langle n|\delta (E-\hat{H})|n\rangle  \nonumber \\
&=& \frac{\partial }{\partial E} \sum_{n} \langle n|\Theta (E-\hat{H})|n\rangle
:= \frac{\partial \Omega (E,\epsilon t)}{\partial E};
\end{eqnarray}
also, we define an energy average over the "frozen" Hamiltonian by
\begin{equation}
\frac{1}{\Sigma }\sum_{n} \langle n|\delta (E-\hat{H}) ... |n\rangle  :=
\langle ...\rangle _{E,\epsilon t},
\end{equation}
which implies
\begin{equation}
\sum_{n} \langle n|...|n\rangle  = \int dE \Sigma \langle ...\rangle _{E,\epsilon t}.
\end{equation}
Now, (18) becomes
\begin{equation}
\Sigma \left( \frac{\partial \hat{\rho }_0'}{\partial E}
\left< \frac{\partial \hat{H}}{\partial (\epsilon t)}
\right>_{E,\epsilon t} +
\frac{\partial \hat{\rho }_0'}{\partial (\epsilon t)}\right) = 0.
\end{equation}
Calling
\begin{equation}
\left< \frac{\partial \hat{H}}{\partial (\epsilon t)} \right>_{E,\epsilon t}
:= u(E,\epsilon t),
\end{equation}
and using (19), we obtain the identity (see Appendix A),
\begin{equation}
\frac{\partial \Sigma }{\partial (\epsilon t)} + \frac{\partial }{\partial E}(\Sigma u) = 0,
\end{equation}
which is a new derivation of quantum adiabatic theorem.
Therefore, (22) reduces to
\begin{equation}
\frac{\partial }{\partial (\epsilon t)}(\hat{\rho }_0'\Sigma) +
\frac{\partial }{\partial E}(u\Sigma \hat{\rho }_0') = 0.
\end{equation}
Then, for $\hat{\rho }_0$, we have
\begin{equation}
\frac{\partial \hat{\rho }_0}{\partial (\epsilon t)}(\{|n\rangle \},\epsilon t) =
\frac{\partial \hat{\rho }_0'}{\partial E}(H,\epsilon t)
\left( \frac{\partial H}{\partial (\epsilon t)} - u \right).
\end{equation}
With this equation and the initial condition (13),
we have completely determined $\hat{\rho }_0$.

The notation $\partial _x$ stands for a partial derivative with respect
to $x$.

We now proceed to determine $\hat{\rho }_1$. The formal
solution of (15) with $j=1$ is
\begin{eqnarray}
\hat{\rho }_1(\{ |n\rangle \},t,\epsilon t) &=& \hat{\rho }_{1i} +
\hat{\rho }_{1h} \nonumber \\
&=& - \int_{0}^{t} dt'
\frac{\partial \hat{\rho }_0}{\partial (\epsilon t)}(\{ |N\rangle  \},\epsilon t)
+ \hat{\rho }_1'(H(\epsilon t), \epsilon t),
\end{eqnarray}
where $|N\rangle  = |N\rangle (|n\rangle ,t,t',\epsilon t)$ is a state reached at time $t'$
evolving backward the state $|n\rangle $ at a time, $t$, under
$\hat{H}(\epsilon t)$. To determine $\hat{\rho }_1'$, we remove
secularities at $O(\epsilon ^2)$ by  a similar procedure as above,
resulting in
\begin{equation}
\frac{\partial }{\partial t}\sum_{n} \langle n|g\hat{\rho }_2|n\rangle  =
-\sum_{n} \left<n\vline g\frac{\partial \hat{\rho }_1}{\partial (\epsilon t)} \vline n\right>,
\end{equation}
or, more explicitly,
\begin{eqnarray}
&&\int dE \frac{\partial }{\partial t} \sum_{n}\langle n|\delta (E-\hat{H})
g\hat{\rho }_2|n\rangle =-\int dE\sum_{n} \left< n\vline \delta(E-\hat{H})
g \frac{\partial \hat{\rho }_1}{\partial (\epsilon t)} \vline n \right> \nonumber \\
&& = \int dE \sum_{n} \langle  n|\delta (E-\hat{H})
g\partial _{{\small \epsilon t}} \int^t dt'
\partial _{{\small \epsilon t}}\hat{\rho }_0(\{|N\rangle \},\epsilon t)|n\rangle
\nonumber \\
&& -\int dE \sum_{n} \langle n|\delta (E-\hat{H})g\partial _{{\small \epsilon t}}
\hat{\rho }_1'|n\rangle
:= T_1 + T_2.
\end{eqnarray}
$T_1$ and $T_2$ are the abbreviations of the two terms above that line
in (29).
Using (26), $T_1$ can be written as
\begin{equation}
T_1 = \int dE \sum_n \langle n|g\delta (E-\hat{H})\partial _{{\small \epsilon t}}
\int_{0}^{t}dt'\partial _{{\small E}}\hat{\rho }_0'
\left( \partial _{{\small \epsilon t}}\hat{H} - u \right)|n\rangle .
\end{equation}
We now employ the notion of distributional or weak derivative of
distributions (denoted by $\delta '(x)$ in the case od delta distributions).
Employing the following property of the Dirac delta distributions
\cite{ls}, viz.,
\begin{equation}
\delta (E-\hat{H})\partial _{{\small \epsilon t}}\Phi =
\delta '(E-\hat{H})\partial _{{\small \epsilon t}}\hat{H} \Phi ,
\end{equation}
(30) becomes
\begin{eqnarray}
T_1&=& \int dE \sum_{n} \langle n|g\delta '(E-\hat{H})
\partial _{{\small \epsilon t}}\hat{H}\int_{0}^{t} dt'
\partial _{{\small E}}\hat{\rho }_0'(\partial _{{\small \epsilon t}}
\hat{H}-u)|n\rangle  \nonumber \\
&=& - \int dE \sum_{n} \langle n|g\delta (E-\hat{H})\partial _{{\small E}}^2
\hat{\rho }_0' \partial _{{\small \epsilon t}}\hat{H}
\int_{0}^{t} dt'(\partial _{{\small \epsilon t}}\hat{H}-u)|n\rangle \nonumber \\
&-&  \int dE \sum_{n} \langle n|g\delta (E-\hat{H})\partial _{{\small E}}
\hat{\rho }_0' \partial _{{\small \epsilon t}}\hat{H}
\int_{0}^{t} dt'\partial _E(\partial _{{\small \epsilon t}}\hat{H}-u)|n\rangle \nonumber \\
\end{eqnarray}
where use has been made of
\begin{equation}
\delta '(E-\hat{H})\partial _{{\small E}}\hat{\rho }_0' =
-\delta (E-\hat{H})\partial _{{\small E}}^2\hat{\rho }_0'.
\end{equation}
This complicated set of terms can be simplified somewhat. To do so, we
rewrite $T_1$ in a way that will help us in introducing two-time
correlation functions later. So,
\begin{eqnarray}
T_1 &=& - \int dE \sum_{n} \langle n|g\delta (E-\hat{H})\partial _{{\small E}}^2
\hat{\rho }_0'
\int_{-t}^{0} ds (\partial _{{\small \epsilon t}}\hat{H}(\{|n\rangle \})-u)
(\partial _{{\small \epsilon t}}\hat{H}(\{|n\rangle \})-u)|n\rangle \nonumber \\
&-&  \int dE \sum_{n} \langle n|g\delta (E-\hat{H})\partial _{{\small E}}
\hat{\rho }_0'
\int_{-t}^{0} ds (\partial _{{\small \epsilon t}}\hat{H}(\{|n\rangle \})-u)
\partial _E(\partial _{{\small \epsilon t}}\hat{H}(\{|N\rangle \})
-u)|n\rangle \nonumber \\
&-&  \int dE \sum_{n} \langle n|g\delta (E-\hat{H})\partial _{{\small E}}
\hat{\rho }_0'
\int_{-t}^{0} ds~~ u
\partial _E(\partial _{{\small \epsilon t}}\hat{H}-u)|n\rangle \nonumber \\
\end{eqnarray}
The average two-time correlation function can now be introduced :
\begin{eqnarray}
&&C_{\epsilon t}(s,E)=
\langle \{ \partial _{{\small \epsilon t}}\hat{H}(\{|n\rangle \},\epsilon t)
-u\} \{ \partial _{{\small \epsilon t}}\hat{H}(\{|N\rangle \},\epsilon t)-u\}
\rangle _{{\small E,\epsilon t}} \nonumber \\
&&= \frac{1}{\Sigma }\sum_{n} \left<n|\delta (E-\hat{H})\{
\partial _{{\small \epsilon t}}\hat{H}(\{|n\rangle \},\epsilon t)-u\}
\{ \partial _{{\small \epsilon t}}\hat{H}(\{|N\rangle \},\epsilon t)-u\}|n\right>.
\end{eqnarray}
Before getting back to (34), we note some simple relations. First of all,
we can write :
\begin{eqnarray}
&~&\Sigma \partial _{{\small E}}\int_{-t}^{0} dsC(s) \nonumber \\
&=&  \Sigma \partial _{{\small E}}\left[\int_{-t}^{0} ds
\frac{1}{\Sigma }\sum_{n} \left< n| \delta (E-\hat{H})\{
\partial _{{\small \epsilon t}}\hat{H}(\{|n\rangle \},\epsilon t)-u\} \{
\partial _{{\small \epsilon t}}\hat{H}(\{|N\rangle \},\epsilon t)-u\}|n\right>
\right]\nonumber \\
&=& -\frac{1}{\Sigma }\frac{\partial \Sigma }{\partial E}
\int_{-t}^{0}ds\sum_{n} \left< n\vline \delta (E-\hat{H})\{
\partial _{{\small \epsilon t}}\hat{H}(\{|n\rangle \})-u\}
\{ \partial _{{\small \epsilon t}}\hat{H}(\{|N\rangle \})-u\}\vline n\right>
\end{eqnarray}
Secondly,
\begin{equation}
\frac{\partial }{\partial E} \left( \Sigma \int_{-t}^{0} ds C(s)\right) = 0.
\end{equation}
With (36) and (37), we can now write a relation,
\begin{equation}
\Sigma \frac{\partial ^2\hat{\rho }_0'}{\partial E^2}\int_{-t}^{0}
ds C(s) = \frac{\partial }{\partial E}\left( \Sigma
\frac{\partial \hat{\rho }_0'}{\partial E}\int_{-t}^{0} ds C(s)\right)
\end{equation}
which we shall use shortly.

To simplify $T_{1}$ and $T_{2}$, we have to employ some further
averaging procedure. We call this a  coarse-graining in which
we replace a function of an eigenvalue, $E_n$ by some average quantity
such that the explicit dependence on the label $n$ disappears. One of the
ways it may be done is by an integration over the average density of states.
The essential point about averaging is that the
spectrum ``seen" by the system is a continuous one.

After some tedious manipulations, repeated usage of the properties of
distributions \cite{ls}, and effecting coarse-graining, we arrive at
\begin{eqnarray}
T_1&=& -2\int dE \overline{g}(E)\overline{\partial ^2_E\hat{\rho }'_{nn}}\Sigma
\int _{-t}^{0} ds C_{\epsilon t}(s,E) \nonumber \\
&-& \int dE \overline{g}(E)\overline{\partial _E\hat{\rho }'_{nn}}\Sigma
\partial _E\int _{-t}^{0} ds C_{\epsilon t}(s,E) \nonumber \\
&+& \int dE \overline{g}(E)\overline{\partial _E\hat{\rho }'_{nn}}
\langle n|\delta (E-\hat{H})\partial _E(\partial _{\epsilon t}H(\{|n\rangle \}))-u).\nonumber \\
&~&.\int_{-t}^{0}ds (\partial _{\epsilon t}H(\{|N\rangle \}))-u)|n\rangle \nonumber \\
&-& \int dE \overline{g}(E)\overline{\partial _E \hat{\rho }'_{nn}}
~u~\langle n|\delta (E-\hat{H}) \nonumber \\
&~&\partial _E \left[
\int_{-t}^{0}ds (\partial _{\epsilon t}H(\{|N\rangle \}))-u)\right]|n\rangle ,
\end{eqnarray}
where we have  employed {\it  ad hoc} coarse-graining and replaced
\begin{eqnarray}
&& g(E_n) ~~\mbox{by} ~~ \overline{g} (E), ~\mbox{and}\nonumber \\
&& \left<n\vline \frac{\partial ^2\hat{\rho }_0'}{\partial E^2}\vline n\right>
~~ \mbox{by} ~~ \overline{\frac{\partial ^2\hat{\rho }_0'}{\partial E^2}}.
\end{eqnarray}
The last two terms are of the same order and opposite sign, so they
will simply compensate for each other.

Because, for times of ${\cal O}(\frac{1}{\epsilon })$,
\begin{equation}
\int_{-t}^{0} ds C(s) = \frac{1}{2} \int_{-\infty }^{\infty } ds C(s)
:= \frac{1}{2} G_2,
\end{equation}
we can write for $T_1$ :
\begin{equation}
T_1 = - \int dE \overline{g} (E)\frac{\partial }{\partial E}
\left[ \Sigma \overline{\frac{\partial \hat{\rho }_0'}{\partial E}}
G_2 \right] - \frac{1}{2}\int dE \overline{g} (E)\Sigma
\overline{\frac{\partial \hat{\rho }_0'}{\partial E}}
\frac{\partial G_2}{\partial E}.
\end{equation}
Notice that $T_2$ has the same form as the expression involving
$\hat{\rho }_0'$ in (18), manipulations are identical. Finally, the
condition that removes secularities to ${\cal O}(\epsilon ^2)$ is
\begin{equation}
\frac{\partial }{\partial (\epsilon t)}(\hat{\rho }_1'\Sigma ) +
\frac{\partial }{\partial E}(u\hat{\rho }_1'\Sigma ) -
\frac{\partial }{\partial E}\left( \Sigma G_2 \overline{\frac{\partial
\hat{\rho }_0'}{\partial E}} \right) - \frac{1}{2}\Sigma
\overline{\frac{\partial \hat{\rho }_0'}{\partial E}}\frac{\partial G_2}{\partial E}=0.
\end{equation}

In terms of occupation probabilities, $p_0$ and $p_1$ (Cf. Eq. (3)), we have
\begin{eqnarray}
&&\frac{\partial }{\partial (\epsilon t)}(p_0\Sigma )
+ \frac{\partial }{\partial E}(up_0\Sigma )=0,  \\
&& \frac{\partial }{\partial (\epsilon t)}(p_1\Sigma )
+ \frac{\partial }{\partial E}(up_1\Sigma )-
\frac{\partial }{\partial E}\left( \Sigma G_2
\frac{\partial p_0}{\partial E} \right)
- \frac{1}{2}\Sigma \frac{\partial p_0}{\partial E}
\frac{\partial G_2}{\partial E} = 0.
\end{eqnarray}
The energy distribution, defined as
\begin{equation}
\eta = \Sigma \left< \hat{\rho }\right>_{E, \epsilon t}
\rightarrow \Sigma \left< \overline{\hat{\rho }} \right>_{E, \epsilon t},
\end{equation}
follows the following equation,
\begin{equation}
\frac{\partial \eta }{\partial t} = -\epsilon
\frac{\partial }{\partial E}(u\eta ) +
\epsilon ^2\frac{\partial }{\partial E}\left[
G_2\Sigma \frac{\partial }{\partial E}\left( \frac{\eta }{\Sigma }
\right)\right]
+ \frac{\epsilon ^2}{2} \Sigma \frac{\partial G_2}{\partial E}
\frac{\partial }{\partial E}\left( \frac{\eta }{\Sigma }\right),
\end{equation}
which is the final result. This equation is different insofar as there
is an extra term as compared to the Smoluchowski equation. Thus, the
diffusion in quantum systems has to be qualitatively
and quantitatively different as the
diffusion coefficient will be different from the one we have in the
Smoluchowski equation.

It is clear that the difference between (47) and the Smoluchowski equation
is the derivative of tegrated time correlation function. This is, indeed,
reminiscent of the relations between friction and diffusion coefficients in
weak-turbulence plasma theory \cite{wallace}. This brings us to the premise
on which we began, the time scales.

First of all, the time scale associated with the decay of correlation function,
\begin{equation}
t_c := [C(0)]^{-1}\int_{-\infty }^{+\infty } C(s)ds.
\end{equation}
If the quantum system considered is modelled as a random matrix of dimension
$N$ \cite{gjv} with large $N$, we know that correlation function will decay
very rapidly. Thus, $t_c$ can be very small if the quantum systems possess the
following properties : (a) the number of eigenvalues is very large, and the
energy spectrum is complex, and, (b) the corresponding classical system is
chaotic. Chaos in the underlying classical system plays a fundamental role
in the decay of correlation functions. It was recently shown \cite{pg-sj} that
the quantum time-depende correlations in a Fermionic system are dominated by the
classical correlation function. The decay of the correlation function is shown
in this work to be governed by the eigenvalues of the Liouvillian operator.
Thus, $t_c $ is related to the Liapunov exponents and other detailed features
of chaos. In classical ergodic adiabatic systems, the time $t$ (fast scale)
is much larger than $t_c$, thus the third term of (47) is zero. However, in
quantal systems, we have the quantum mechanical scale, $t_q=\hbar /S$
($S$ being the mean level spacing) which is why
the third term at ${\cal O}(\epsilon ^2)$ is explicitly present.
If $t_q \ll t_c \ll t$, the quantum effects will dominate, and all the terms
in (47) will be important. If $t_c \ll t_q \ll t$, then the system will
behave classically initially and eventually, quantum phenomena will become
important; so initial evolution will be Smoluchowski-like and then
non-Smoluchowski regime sets in. If, however, $t_c \ll t \ll t_q$, then the
evolution will be according to the classical equation. Notice, as $\hbar $
becomes small and the system is classical, $t_q$ will become large, which
explains how (47) will reduce to the Smoluchowski equation.

In this paper, we have given a formal proof, which is important for any
new equation. It is our belief that examples will help in understanding
(47) more. We wish to emphasise that the derivation does not assume anything
special about the initial density operator (like, e.g., the Kubo-Martin-
Schwinger condition). This generality is very important to note. However,
we have employed {\it ad hoc} coarse-graining which, hopefully, does not
destroy the novelty.
\newpage
\begin{center}
{\bf Acknowledgements}
\end{center}

For a part of this work,
the author was  financially supported by the "Communaute Francaise de Belgique"
under contract no. ARC-93/98-166 on ``Quantum keys to reactivity".
It gives him  great pleasure to thank
Pierre Gaspard, Chris Jarzynski,  and Mitsusada Sano for
some interesting
discussions at different stages of this work.
\newpage
\begin{center}
{\bf APPENDIX - A}
\end{center}
To prove (24), observe that
\begin{eqnarray}
\frac{\partial }{\partial E}(\Sigma u) &=& \frac{\partial }{\partial E}
\left[ \sum_{m} \langle m|\delta (E - \hat{H})|m\rangle \frac{1}{\Sigma }
\sum_{n} \langle n|\partial _{\epsilon t}\hat{H}\delta (E - \hat{H})|n\rangle
\right] \nonumber \\
&=& \frac{\partial }{\partial E}
\sum_{n} \langle n|\partial _{\epsilon t}\hat{H}\delta (E - \hat{H})|n\rangle
\nonumber \\
&=& \sum_{n} \langle n|\partial _{\epsilon t}\hat{H}\delta ^{\prime}
(E - \hat{H})|n\rangle \nonumber \\
&=& - \frac{\partial }{\partial (\epsilon t)}
\sum_{n} \langle n|\delta (E - \hat{H})|n\rangle \nonumber \\
&=& - \frac{\partial \Sigma }{\partial (\epsilon t)}.
\end{eqnarray}
The last equality follows because $\langle n | m \rangle = \delta _{nm}$
implies that
\begin{eqnarray}
&&\left( \frac{\partial }{\partial (\epsilon t)}\langle n|\right) \delta
(E - \hat{H})|n\rangle + \langle n|\delta (E - \hat{H})
\left( \frac{\partial }{\partial (\epsilon t)}|n\rangle \right) \nonumber \\
&&= \langle \dot{n}|\delta (E - \hat{H})|n\rangle +
\langle n|\delta (E - \hat{H})|\dot{n}\rangle \nonumber \\
&&= [\langle \dot{n}|n\rangle + \langle n|\dot{n}\rangle ]\delta (E-E_n) = 0.
\end{eqnarray}

\end{document}